\documentclass{pasa}%

\title[Spectral calibration of interferometers for EoR]{Spectral calibration requirements of radio interferometers for Epoch of Reionisation science with the SKA}
\author[Trott \& Wayth]{Cathryn M. Trott$^{1,2}$ \and Randall B. Wayth$^{1,2}$\thanks{email: cathryn.trott@curtin.edu.au}\\
\affil{$^1$International Centre for Radio Astronomy Research, Curtin University, Bentley WA 6103 Australia}%
\affil{$^2$ARC Centre of Excellence for All-Sky Astrophysics (CAASTRO)}}%
\jid{PASA}
\doi{10.1017/pas.\the\year.xxx}
\jyear{\the\year}

\usepackage[authoryear]{natbib}
\bibpunct{(}{)}{;}{a}{}{,}
\usepackage{aas_macros}
\usepackage{hyperref,graphicx,subfigure} 
\hypersetup{colorlinks,citecolor=blue,linkcolor=blue,urlcolor=blue}

\begin{document}%
\begin{abstract}
Spectral features introduced by instrumental chromaticity of radio interferometers have the potential to negatively impact the ability to perform Epoch of Reionisation (EoR) and Cosmic Dawn (CD) science using the redshifted neutral hydrogen emission line from the early Universe. We describe instrument calibration choices that influence the spectral characteristics of the science data, and assess their impact on EoR statistical and tomographic experiments. Principally, we consider the intrinsic spectral response of the receiving antennas, embedded within a complete frequency-dependent primary beam response, and frequency-dependent instrument sampling. We assess different options for bandpass calibration.
The analysis is applied to the proposed SKA1-Low EoR/CD experiments.
We provide tolerances on the smoothness of the SKA station primary beam bandpass, to meet the scientific goals of statistical and tomographic (imaging) EoR programs.
Two calibration strategies are tested: (1) fitting of each fine channel independently, and (2) fitting of an $n$th-order polynomial for each $\sim$1~MHz coarse channel with ($n$+1)th-order residuals ($n$=2,3,4).
Strategy (1) leads to uncorrelated power in the 2D power spectrum proportional to the thermal noise power, thereby reducing the overall array sensitivity.
Strategy (2) leads to correlated residuals from the fitting, and residual signal power with ($n$+1)th-order curvature. For the residual power to be less than the thermal noise, the fractional amplitude of a fourth-order term in the bandpass across a single coarse channel must be $<$2.5\% (50~MHz), $<$0.5\% (150~MHz), $<$0.8\% (200~MHz). The tomographic experiment places stringent constraints on phase residuals in the bandpass. We find that the root-mean-square variability over all stations of the change in phase across any fine channel (4.578~kHz) should not exceed 0.2 degrees.
\end{abstract}
\maketitle%
\section{Introduction }
\label{sec:intro}
Detection of the signal from the Epoch of Reionisation (EoR), and estimation of the spatial properties of neutral hydrogen from the early Universe, are challenging experiments, requiring high fidelity data with known statistical properties \citep{koopmans15,parsons10,vanhaarlem13,bowman13_mwascience}. For these, instrumental effects that bias the signal can have a large impact on the success of the experiment, and interpretation of results. Statistical experiments, such as measurement of the spatial power spectrum of neutral hydrogen (HI) brightness temperature fluctuations, implicitly rely on datasets that have been observed with a spectrally-smooth instrument, in order to associate spectral structure with intrinsic spatial fluctuations. Similarly, HI tomography (imaging as a function of fine spectral channel) aims to detect and measure weak spectral lines from HI clouds, thereby requiring smooth instrumental spectral characteristics in order to not bias the signal. Further, phase residuals in the calibration of stations reduces imaging dynamic range, destroying the detectability of weak signals \citep{perley99}.

Low-frequency aperture array interferometers, such as the Square Kilometre Array (SKA-Low), use the electronic combination of individual element antennas in an `aperture array', to form a single beam on the sky. This set of beamformed elements forms a single station. These stations are then used as the primary aperture with which to form cross-correlations for interferometric visibilities.
The advantages of aperture array telescopes include flexible and dynamic allocation of elements to a station, flexible and dynamic assignment of weights to alter the beam shape (apodisation), and durability due to the lack of moving parts. Such trade-offs are being studied for the SKA \citep{grainge14,mort16}.
The disadvantages include direction-dependent primary beam shapes, strong off-zenith instrumental polarisation, and mutual coupling between elements \citep{sutinjo15}.

Calibration of radio interferometers is crucial to obtaining science-quality data.
For precision quantitative experiments, accurate and precise calibration is required for reliable results to be achieved and claimed. Calibration is the process by which the electronic gain amplitude and phase is set for each receiving element, such that the sky position and source flux density of any sky source is correct.
Calibration of interferometric arrays implies measurement and setting of the complex gain parameters for each station, as constructed from the gains of each element.
Direction-independent, but frequency-dependent bandpass and phase calibration is required for each station that will form an interferometric baseline. %Typically, for small field-of-view instruments, at frequencies where a single bright point source, with known flux density, in the sky dominates the received power, the instrument is calibrated by `phasing' the cross-correlation products (visibilities) to the bright point source and setting the gains such that the station phases are zero, and the amplitudes correspond to the catalogue flux density of the source. The gain calibration procedure estimates $2N_{\rm ant}F$ parameters using $N_{\rm ant}(N_{\rm ant}-1)F$ measurements, where $N_{\rm ant}$ is the number of stations and $F$ is the number of frequency channels. For wide field-of-view instruments, where the sky background, or multiple point sources contribute to the received power, the concept is the same, but a more sophisticated input sky model is required. This is the case for low-frequency interferometers, where the long wavelength and increased sky temperature demand an accurate input model. Further, at these frequencies, differential refraction due to ionospheric plasma leads to spatial and time-dependent changes to the static sky, requiring short timescale \citep[$\sim$minutes,][]{loi15} re-calibration of the gain parameters. Such high cadence measurements limit the amount of information usable with which to perform the gain measurements, thereby setting an upper limit to the precision with which they can be measured.
The calibration process for a low-frequency interferometer typically involves simultaneously solving for station direction-independent and -dependent complex-valued gains (possibly including ionospheric effects) using a sky model and primary beam model \citep{mitchell08,yatawatta08,kazemi13,tasse13}.
Principally, each antenna will have a bandpass that must be calibrated.
The simplest model is to treat each frequency channel as being independent, but other approaches can be used if characteristics of the bandpass shape are known. In general, we are separating the direction-independent bandpass fitting from the direction-dependent beam fitting, but these may be combined to exploit the expected characteristics of the instrument \citep[e.g., ][where a smooth polynomial is used to regularise the bandpass solution in a full direction-dependent calibration]{yatawatta15}. In these more advanced approaches, instrument responses that fail the underlying smoothness tests (e.g., rapid changes in the bandpass shape) will still leave residual bandpass structure, and yield potential biases in the final calibrated data. Here, we try to take a general approach to fitting of the bandpass, under the assumption of a smooth response in frequency.

Understanding the instrument spectral response is especially important for EoR and Cosmic Dawn experiments \citep{trottchips15,offringa16,barry16}. Probing an emission line over cosmological volume is achieved by detecting and measuring the line as a function of redshift, and hence as a function of frequency.
For statistical experiments where the spatial structure of HI is probed by a Fourier-like Transform along frequency (obtaining information about the characteristic size scale of neutral and ionised regions), unmodelled instrumental spectral structure that is localised in frequency space contaminates the entire signal space after Fourier transform, thereby biasing results.
Likewise, for imaging experiments, which aim to directly detect and map individual ionised bubbles, spectral structure can mimic these structures. Further, residual phase in the data smears the signal from other sources in the field, reducing the contrast (dynamic range) between the background and the bubble of interest.

In addition to the \textit{imprecision} of gain calibration parameters due to limited information, there is the potential for \textit{inaccurate} calibration if an input sky model, or input instrument model is incomplete or incorrect.
%The effect of the instrument on the signal is felt through both the accurate understanding of the primary beam response to the sky at a given sky position and at a given frequency.
The latter is determined by the bandpass shape of the station antennas and electronics, and may be element- and time-dependent.
%Improperly modelling this bandpass shape can lead to spectral residuals in the data. %Moreover, calibration schemes that fail to reproduce any structure in the bandpass can also leave residuals in the signal.
Early testing of the SKA Low-frequency Aperture array element dipoles (SKALA) show strong spectral features \citep{delera15}.
%In addition, these features are time-dependent, due to their dependence on the shape of the dipole (affected by the wind) and ambient temperature.
These features have the potential to leave residual signal in the data if calibration is not carefully performed. In this work, we describe two common calibration methods, and explore the effects of imprecise and inaccurate calibration for SKA-Low for the planned EoR experiments.

\section{Experiments and their science requirements}
The EoR/CD program includes three major experiments:
\begin{enumerate}
\item Power spectrum: The underlying data for the power spectrum estimation will be visibilities averaged to $\sim$100~kHz and 5~seconds. Simplistically, these visibilities are gridded and integrated, and then squared and normalised by volume to form a power spectrum (units: K$^2$ Mpc$^3$). Tolerances for spectral gradients in the post-calibration bandpass will be computed on this basis, using the station calibration timescale to define the information available for calibration;
\item Tomography: The imaging data consist of visibilities averaged to $\sim$100~kHz and integrated; %The relevant timescale is the calibration timescale;
\item 21~cm Forest: The data will be high spectral resolution image cubes, with fine spectral channels ($\sim$4~kHz).
\end{enumerate}
%The power spectrum (statistical) experiments, across the frequencies of interest to EoR, will be used as the primary assessment metric.
Table \ref{table:experiments} lists the relevant parameters of the data and experiments, as used in this analysis.
\begin{table*}[ht]
\centering
\begin{tabular}{|l|l|l|l|l|l|}
\hline 
Experiment & $\Delta\nu$ &  $\Delta{t}$ & Product & Total time & $\Delta_T^2$ ($k$=0.03~Mpc$^{-1}$) \\ 
\hline
Power spectrum (54~MHz) & 109.8~kHz$^1$ & 5~s & Visibilities & 1000~h & 0.2~mK$^2$\\
Power spectrum (84~MHz) & 109.8~kHz$^1$ & 5~s & Visibilities & 1000~h & 0.02~mK$^2$\\
Power spectrum (143~MHz) & 109.8~kHz$^1$ & 5~s & Visibilities & 1000~h & 0.005~mK$^2$\\
\hline
Experiment & $\Delta\nu$ &  $\Delta{t}$ & Product & Total time & $\Delta{T}$\\
\hline
Tomography ($>$50~MHz) & 100~kHz & 1000~h & Image cube & 1000~h & 1~mK\\
\hline
Experiment & $\Delta\nu$ &  $\Delta{t}$ & Product & Total time & $\Delta{S}$\\
\hline
21~cm absorption ($>$50~MHz) & 4.5~kHz & 1000~h & Image cube & 1000~h & 0.2~mJy\\
\hline
\end{tabular} 
\caption{Spectral and temporal parameters of EoR/CD experiments, as described in \citet{koopmans15}. $^1$Averaging to match an integer number of fine channels.}
\label{table:experiments}
\end{table*}

The current SKA Level 1 requirement for station complex gain calibration is 600~seconds\footnote{SKA System Requirement SYS\_REQ-2635}. This timescale will be used to determine the temporal tolerance of the bandpass stability. The spectral resolution allows 65,536 fine channels across the 300~MHz bandwidth\footnote{SKA System Requirement SYS\_REQ-2148}, yielding 4.578~kHz channels.

\section{Approach}
We use the array baseline design\footnote{Described in Document SKA-SCI-LOW-001 (`SKA1-LOW Configuration')}, incorporating 94 superstations of six logical $\sim$33~m stations arranged in a floral configuration. This configuration provides a large amount of sensitivity to angular modes on the sky corresponding to the $\sim$33~m intra-super-station baseline, and twice this length.

{\bf Power spectrum:} We compute the thermal noise uncertainty for the experiments described in Table \ref{table:experiments}, for the 2D (cylindrical) power spectrum. The $uv$-plane resolution is defined by the field-of-view of the telescope at a given frequency. We choose an experiment bandwidth of 10~MHz around three central frequencies: 50~MHz ($z\sim$~25), 150~MHz ($z\sim$~8.5), 200~MHz ($z\sim$~6.1). This thermal noise estimate provides the reference level, below which, spectral features may be tolerated.

We compare the thermal noise and bias tolerance to the signal in the power spectrum due to calibration errors and residuals, using a statistical model of the point source population. The following approach is taken:
\begin{enumerate}
\item Compute the signal available from the sky to perform calibration, using a statistical model of unresolved point sources, as observed by an interferometer with a frequency-dependent primary beam. This model serves two purposes: (a) As a reference signal of the power in the sky, and its statistical signature in power spectrum parameter space; (b) As the source population to which calibration is applied (i.e., the sky model), and from which residual signal stems after application of an incorrect calibration model;
\item Compute the precision with which calibration is achievable with an ideal estimator for two calibration schemes, based on the statistical sky model, and propagate uncertainties into the power spectrum;
\item Compute the residual signal (accuracy) from the sky model due to unfitted spectral structure in the calibration process, and propagate into the power spectrum.
\end{enumerate}

{\bf Tomography:} We compute the loss in dynamic range in an image cube due to residual phase gradients across the fine frequency channels. We compute the residual gradient in phase that may be tolerated such that an intrinsic 1~mK brightness temperature fluctuation in a 100~kHz spectral channel is detectable.

\section{Power spectrum}
\subsection{Signal due to calibration uncertainties and residual spectral structure}
One possible calibration approach for EoR will be composed of a set of steps to model and remove sky signal, potentially including a direct subtraction of sources through a Global Sky Model (GSM). This step occurs after calibration of each station's bandpass, and assumes accurate and precise calibration. There are two sources of residual spectral signal due to calibration: (1) In the case of bandpass calibration \textit{uncertainties}, residual noise-like sky signal will remain in the data from fitting of the bandpass calibration parameters using the information available in the sky; (2) Residual sky signal power and structure will remain from any spectral curvature terms that are not accounted for in the calibration model. Subtraction of the GSM will leave these residuals in the data due to incomplete and imprecise calibration of the complex station gain parameters.

The model and approach used for bandpass calibration will affect the calibration precision and the residual signal. Conceptually there is a trade-off between precision and complexity of the model. For example, a polynomial fitted in amplitude and phase to each coarse channel of order $N$ will yield a handful of parameters with precision dependent on the information available in the data obtained over the calibration timescale, and leave residual signal with spectral curvature of order $>N$. The parameter estimates are also likely to be correlated in frequency, thereby correlating the fine frequency channels. Conversely, each fine channel can be independently fitted, using a single amplitude and phase estimate. This would yield uncorrelated uncertainties between fine channels, but requires the fitting of more parameters, thereby reducing signal-to-noise.

In this work we explore both precision and accuracy. We compute the signal introduced by imprecise bandpass calibration, leading to additional noise in the data. We then consider gain amplitude and phase residuals, whereby the calibration model is incorrect, and spectral structure remains. The former introduces additional signal due to noise-like fluctuations in the visibilities. The latter introduces additional signal into the power spectrum due to residual signal from the sources in the sky. In all cases described, we are computing signal power that is \textit{additive} to thermal noise and sky power in the power spectrum space.

We take two common calibration approaches --- low-order polynomial fitting and individual fine channel fitting --- and compute the precision with which these parameters can be estimated and their correlations using the Cramer-Rao Bound \citep{kay93}. The parameters of importance are the complex-valued gains for each station (or, equivalently the amplitude and phase) over a solution interval in frequency. The uncertainties on these parameters, which may be correlated, are propagated back into the bandpass solution, and through to the power spectrum, to yield the \textit{calibration uncertainty power}. Remaining spectral curvature is then propagated into the power spectrum to yield the \textit{residual spectral power}. The former sets the noise floor for the calibration method, while the latter informs the spectral gradient tolerance for the bandpass solution by determining how much additional bias signal is introduced into the experiment.

In all cases we are considering real sky signal that remains in the data, and yields signal in the power spectrum that is additive to the thermal noise power.
We also do not specify an actual procedure for performing the calibration.
Rather, the approach is valid for any general, unbiased estimator hence represents the best possible result that can be obtained using the proposed models.
%This implies that our results are valid for any general fitting scheme.

\subsection{Formalism}
We use a statistical model of the instrument, which allows us to propagate spectral signal into the power spectrum parameter space. This model includes all instrumental spectral features, including a frequency-dependent primary beam, and instrumental layout (configuration). The covariance between frequency channels $\nu$ and $\nu^\prime$, at angular wavenumber $u$ is given generically by \citep{trottchips15}:
\begin{eqnarray}
\boldsymbol{C}_{\rm FG}(\nu,\nu^\prime;u) &=& \rho(\nu;\nu^\prime;u)\label{eqn:rho}\\\nonumber &\int_0^\infty& B(l;\nu)B(l;\nu^\prime) J_0\left({2\pi(ul)(\nu-\nu^\prime)}\right) ldl \hspace{0.3cm} {\rm Jy^2}.
\end{eqnarray}
Here, $B(l,m; \nu)$ describes the frequency-dependent primary beam response to the sky at radial position $l=|{\bf l}|$, $\rho$ contains the frequency correlations at spatial wavemode $u$, and the Bessel function performs the Fourier Transform to wavenumber space at mode $u$ ($\propto k_\bot$). This expression is appropriate for a circularly-symmetric primary beam, and allows the propagation of any spectral function into the power spectrum. It is then propagated into the 2D power spectrum parameter space according to a Fourier transform:
\begin{equation}
\langle P_k(k_\bot,\eta) \rangle = \mathcal{F}^\dagger \boldsymbol{C}_{\rm FG} \mathcal{F},
\label{eqn:fg_power}
\end{equation}
with relevant cosmological co-ordinate conversions at a given frequency (redshift) to convert the line-of-sight wavenumber, $\eta$ (Hz$^{-1}$), to $k_\parallel$ (Mpc$^{-1}$), and the angular wavenumber $u$ to $k_\bot$.

\subsubsection{Reference signal power from the sky}
To estimate the power due to astrophysical sources in the sky, we use a statistical model for extragalactic point sources, based on a parametric source counts function and a Poisson random position distribution across the sky. The model used, and the framework for it, are described in \citet{trottchips15} and \citet{trott15}. The number of sources of a given flux density in a unit area of sky is Poisson-distributed ($\mathcal{P}()$):
\begin{equation}
N(S,S+dS) dS\,d{\bf l} \sim \mathcal{P}(\langle{N}\rangle),
\end{equation}
where $\langle{N}\rangle$ is the expected number, given parametrically by:
\begin{eqnarray}
\langle{N(S,S+dS)}\rangle(\nu) &=& \frac{dN}{dS}(\nu)\,dS\,d{\bf l} \\&=& \alpha \left( \frac{\nu}{\nu_0} \right)^{\gamma} \left( \frac{S_{\rm Jy}}{S_0}\right)^{-\beta}\,dS\,d{\bf l}.
\label{source_counts}
\end{eqnarray}
We use values of $\alpha=4100\,{\rm Jy}^{-1}{\rm sr}^{-1}$, $\beta=1.59$ and $\gamma=-0.8$ at 150~MHz, in line with recent measurements \citep{intema11,gervasi08}. This Poisson statistical model is propagated into the power spectrum, using a full frequency-frequency covariance to describe the spectral correlations, and instrument sampling to produce the `wedge' effect of foreground contamination.

The frequency-frequency covariance at angular wavenumber $u=|x|c/\nu_0$, for a Gaussian-shaped primary beam of a station of diameter $D$ is given by \citep{trottchips15}:
\begin{eqnarray}
\boldsymbol{C}_{\rm FG}(\nu,\nu^\prime) &=& \frac{\alpha}{3-\beta} \left( \frac{\nu}{\nu^\prime} \right)^{-\gamma} \frac{S_{\rm max}^{3-\beta}}{S_0^{-\beta}} \frac{\pi{c^2}\epsilon^2}{D^2}\frac{1}{\nu^2 + \nu^{\prime{2}}}\label{eqn:fg_model}\\\nonumber && \exp{\left( \frac{-u^2c^2f(\nu)^2\epsilon^2}{4(\nu^2 + \nu^{\prime{2}})D^2} \right)},
\end{eqnarray}
where $\epsilon=0.42$ converts an Airy disk to a Gaussian characteristic width, and $f(\nu) = (\nu-\nu^\prime)/\nu_0$. This model describes the covariance for a full extragalactic point source population within the field-of-view of an SKA1-Low station. Equation \ref{eqn:fg_model}, propagated according to equation \ref{eqn:fg_power}, then contains the expected signal in the 2D parameter space due to sources within the field-of-view, for a perfectly calibrated instrument. For the SKA, we consider a 12~MHz bandwidth for each experiment (16 coarse channels), and a FOV associated with a 35~m station at each frequency. This yields a power spectrum parameter space (at 150~MHz) of $k_\parallel=[0.01,1.60]h$~Mpc$^{-1}$, $k_\bot=[0.02,2.07]h$~Mpc$^{-1}$.

\begin{figure*}[t]
{
\includegraphics[width=1.1\textwidth,angle=180]{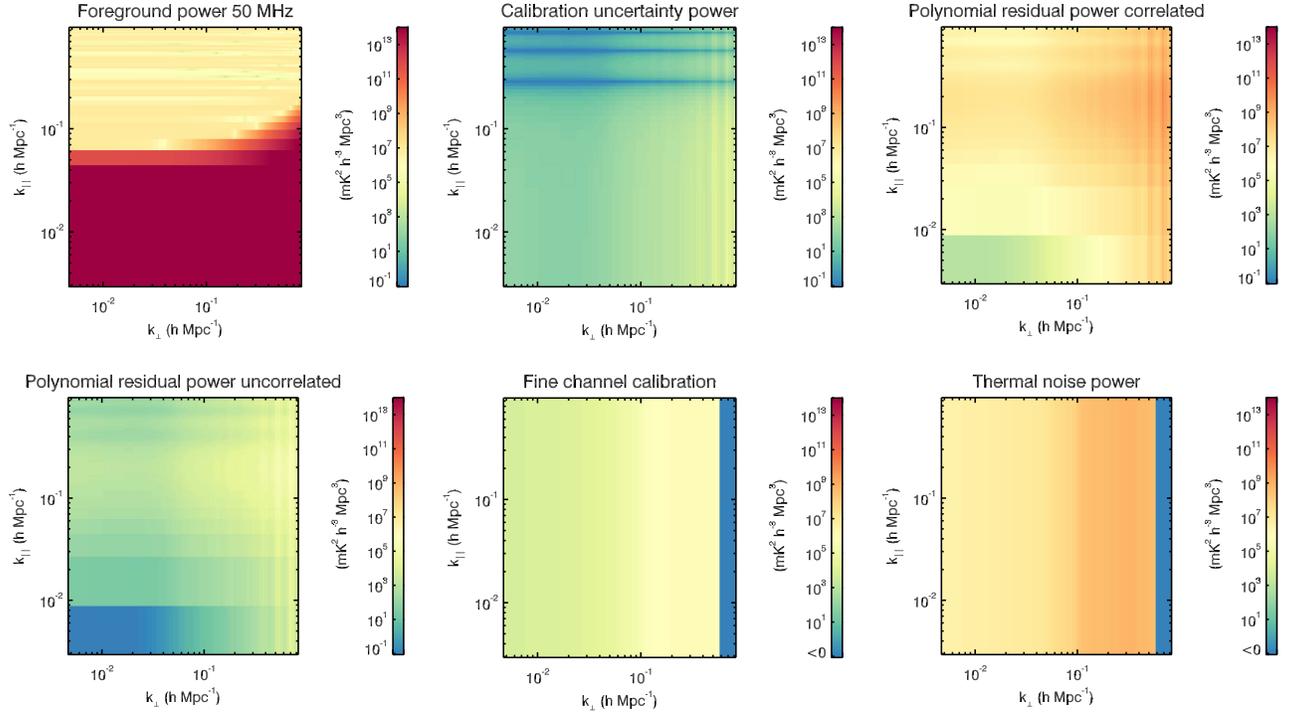}
}
\caption{50~MHz: (Top-left) Reference unresolved point source model, propagated into the power spectrum; (top-centre) the calibration uncertainty power for a third-order polynomial fit; (top-right) residual unresolved point source power in the most pessimistic calibration model (scaled to the tolerance level); (bottom-left) the residual unresolved point source power due to residual fourth-order curvature (most optimistic model); (bottom-centre) the uncertainty power due to calibration of each fine channel independently; (bottom-right) Thermal noise power.}
\label{fig:power_in50}
\end{figure*}

\begin{figure*}[t]
{
\includegraphics[width=1.1\textwidth,angle=180]{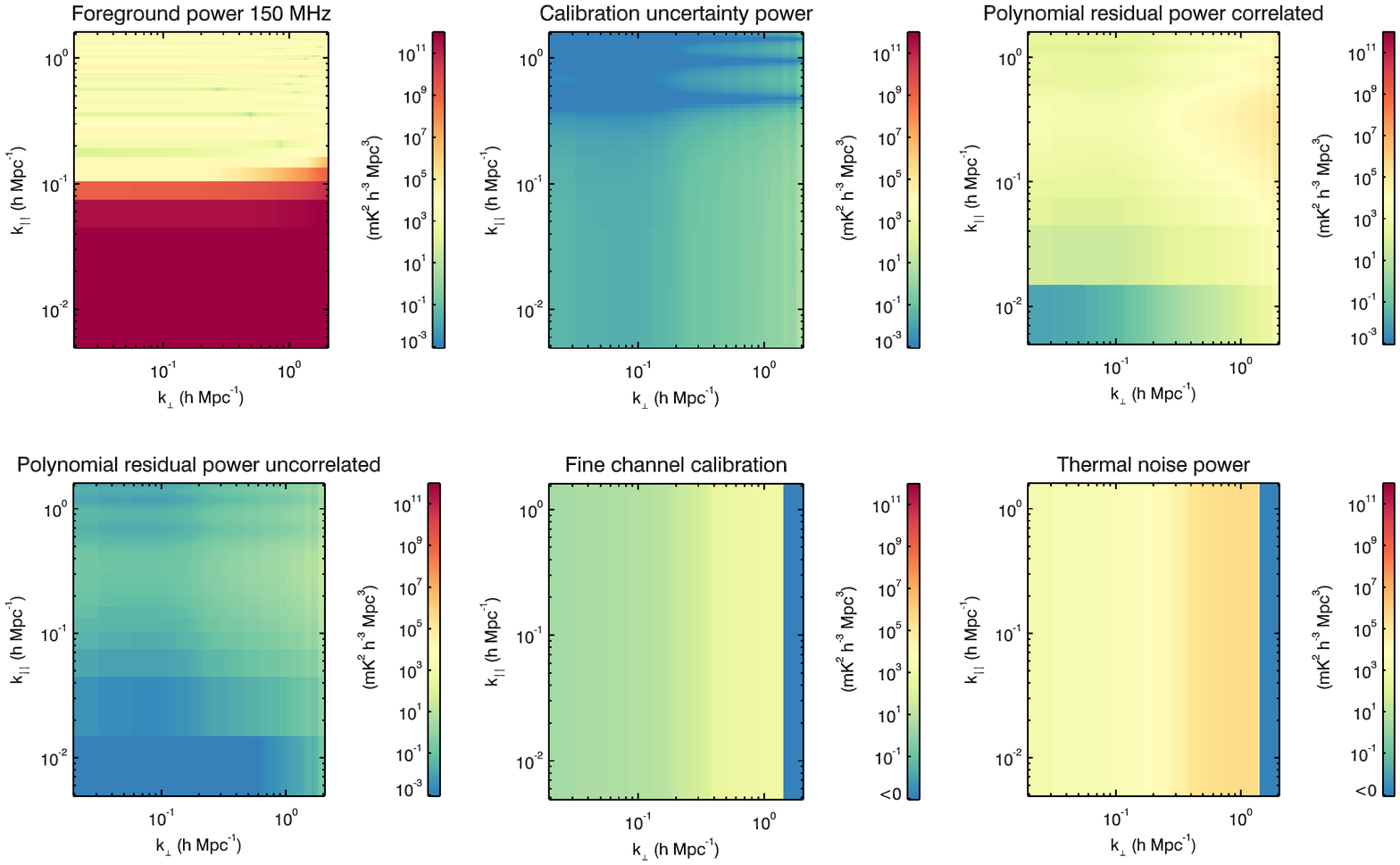}
}
\caption{150~MHz: (Top-left) Reference unresolved point source model, propagated into the power spectrum; (top-centre) the calibration uncertainty power for a third-order polynomial fit; (top-right) residual unresolved point source power in the most pessimistic calibration model (scaled to the tolerance level); (bottom-left) the residual unresolved point source power due to residual fourth-order curvature (most optimistic model); (bottom-centre) the uncertainty power due to calibration of each fine channel independently; (bottom-right) Thermal noise power.}
\label{fig:power_in150}
\end{figure*}

\begin{figure*}[t]
{
\includegraphics[width=1.1\textwidth,angle=180]{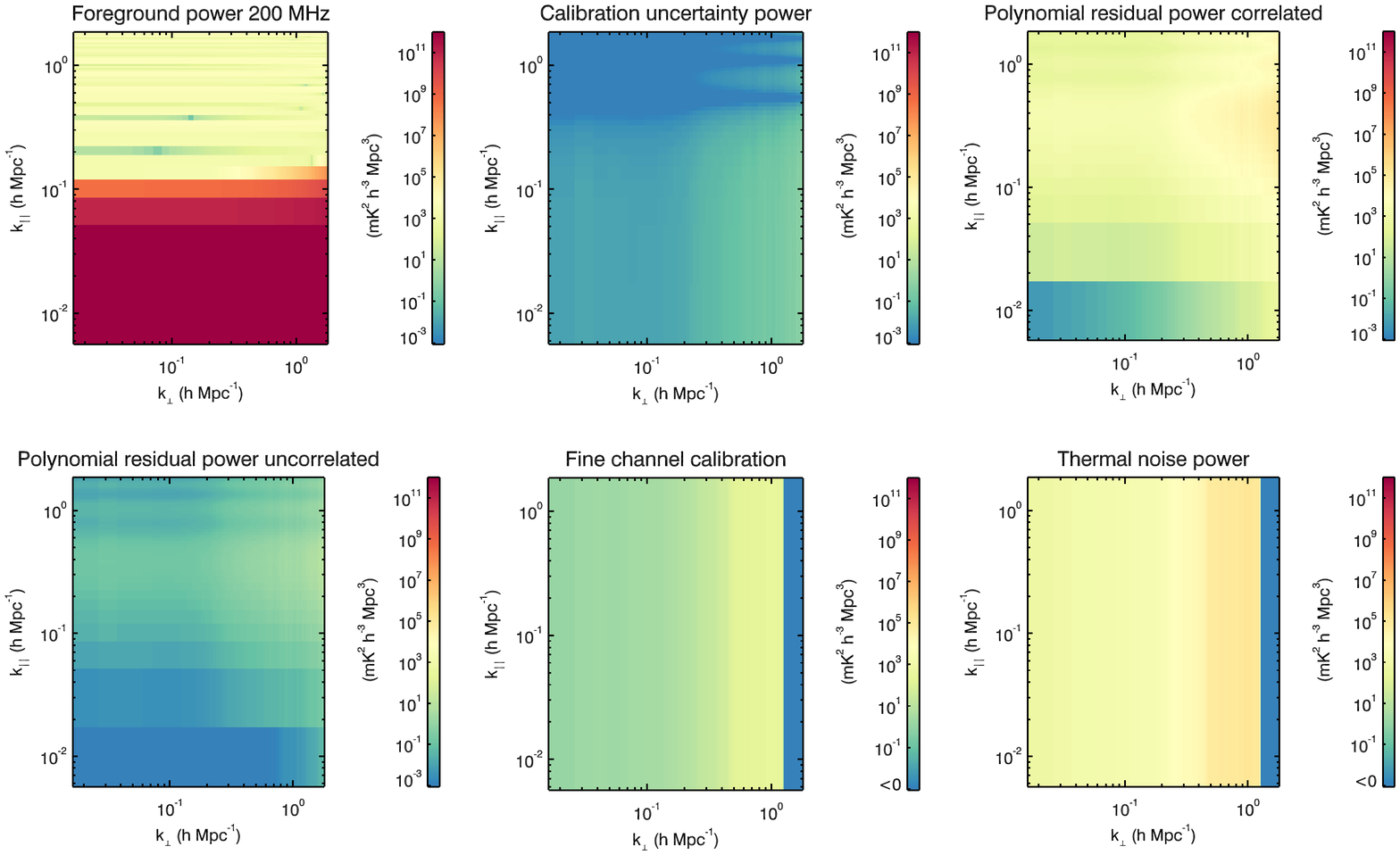}
}
\caption{200~MHz: (Top-left) Reference unresolved point source model, propagated into the power spectrum; (top-centre) the calibration uncertainty power for a third-order polynomial fit; (top-right) residual unresolved point source power in the most pessimistic calibration model (scaled to the tolerance level); (bottom-left) the residual unresolved point source power due to residual fourth-order curvature (most optimistic model); (bottom-centre) the uncertainty power due to calibration of each fine channel independently; (bottom-right) Thermal noise power.}
\label{fig:power_in200}
\end{figure*}

The top-left panels of Figures \ref{fig:power_in50}, \ref{fig:power_in150} and \ref{fig:power_in200} display the power due to unresolved point sources for the experiments described in Table \ref{table:experiments} and a 2D, cylindrically-averaged parameter space ($k_\bot, k_\parallel$).
%The evolution in the thermal noise power with $k_\bot$ reflects the density of baselines for the instrument, with a large number of very short baselines.
All unresolved sources with a flux density below 5~Jy remain in the data. This is to ensure the model is statistically-consistent (i.e., brighter sources are rare and not accurately characterised by this statistical model). The unresolved point source model describes the total sky power as observed by the instrument, with a frequency-dependent primary beam. At its core is a frequency-frequency covariance matrix that describes all the frequency correlations of signal in the data. It is this covariance matrix that can be used to approximate spectral bandpass features, and their effects in the EoR power spectrum. The full reference point source model described above provides the complete signal due to sources in the sky without any signal subtraction.

\subsection{Calibration uncertainty}
Both calibration schemes (polynomial and fine-channel fitting) are expected to have temporally uncorrelated uncertainties between calibration cycles, yielding a factor of $T_{\rm obs}/T_{\rm cal}$~=~1000-h/600-s~=~6000 reduction in calibration uncertainty power over the full experiment. It is often implicitly assumed that individual stations will have uncorrelated \textit{errors}, yielding a stochastic reduction by a factor equivalent to the $uv$-sampling over the 4-hour nightly track (we term this the `optimistic' case). These are taken into account in the modelling. In particular, the estimation of parameters for each station uses visibilities for the whole array, yielding a factor of $(N_{\rm ant}-1)/2$ improvement in power. In the case where short baselines are omitted from the calibration (to avoid large-scale structure biasing the procedure), fewer baselines are available and the improvement will be reduced. For SKA1-Low, there are many short baselines in the core, and within each of the superstations. Of the 158,000 baselines, $>$4,000 are shorter than 50 wavelengths at 150~MHz, effectively reducing the improvement from measurements from a factor of $(N_{\rm ant}-1)/2\simeq{280}$ to $\sim$270. This has, therefore, only a minor impact.

\subsubsection{Polynomial fitting}
\begin{figure}[t]
{ \includegraphics[width=.5\textwidth]{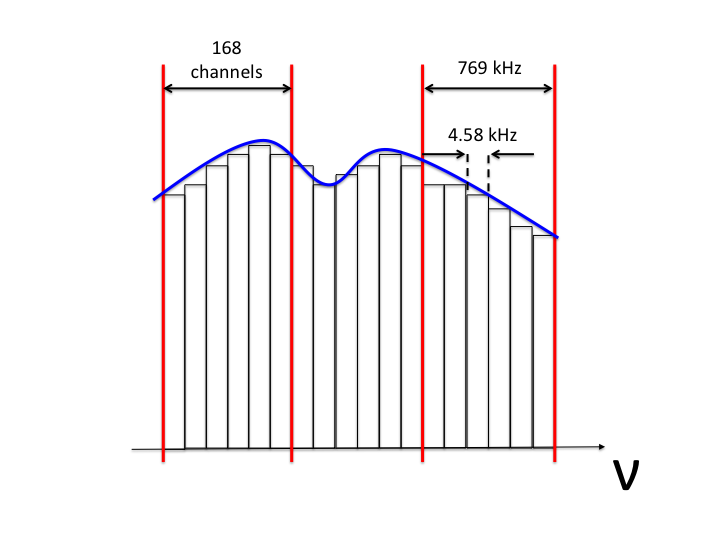}}
\caption{Schematic figure showing a smooth fit over three contiguous coarse channels, where the fit is performed on a fine channel basis. The fitting parameters derived from these three coarse channels are used to calibrate the central channel only. Each vertical bar denotes a single fine channel, and the red lines denote coarse channel band edges.}
\label{fig:bandpass}
\end{figure}

One approach to bandpass calibration assumes that the bandpass may be fitted with a low-order polynomial function. This has the advantage of estimating only a handful of parameters over the bandwidth. In this work, we take a reasonable approach of fitting for each coarse channel independently, but using three contiguous coarse channels to perform the fit. The real and imaginary components are fitted as separate polynomials. We assume that a $n$th-order polynomial ($n$=2,3,4) is fitted over these three coarse channels independently to each of the real and imaginary components of the visibilities, with $N_{\rm fine}=168$ fine channels of $\nu_{\rm fine}=4.578$~kHz within a coarse band ($\Delta\nu_{\rm coarse}=769.043$~kHz). Each fit therefore uses 3*168=504 datapoints, with 600~s of data and the full sky power, and there are two fits (the information for which is contained in twice as many datapoints when counting the real and imaginary components of the complex visibilities). Figure \ref{fig:bandpass} shows this calibration scheme.
For the third-order polynomial, we fit four parameters, $A,B,C,D$, such that:
\begin{equation}
S_{r,i}(\nu) = A(\nu-\nu_0)^3 + B(\nu-\nu_0)^2 + C(\nu-\nu_0) + D,
\end{equation}
where $r,i$ index real and imaginary, and $\nu_0$ is defined as the centre of the coarse channel. The Fisher Information Matrix and the subsequent propagation of the correlated errors back into the signal, are independent of the actual values of $A,B,C$ and $D$ (due to the fact that they are linear), thereby making this a general third-order model. The 2nd and 4th-order fits have one fewer or one more parameter. The amplitude, $\mathcal{A}$, of the calibration solution is given, as usual, by:
\begin{equation}
\mathcal{A}(\nu) = \sqrt{S_r(\nu)^2 + S_i(\nu)^2},
\end{equation}
and the phase, $\Phi$, by:
\begin{equation}
\Phi(\nu) = {\rm atan}\left(\frac{S_i}{S_r} \right).
\end{equation}

\subsubsection{Fine-channel fitting}
An alternative calibration method is to estimate the amplitude of each fine frequency channel independently. This has the advantage of removing any residual correlation between channels, but, requiring estimation of 168 parameters over a single coarse channel, requires sufficient information to be available on the calibration timescale (i.e., sufficient signal-to-noise in 600~s to obtain a precise estimate). The consequent calibration precision will be lower for more parameters, but spectral correlations will be removed, and \textit{no residual curvature will remain, except within a fine channel.}

%Due to the estimation of a single amplitude per fine channel, which is assumed to be independent between stations and over different calibration cycles, and unbiased, \textit{the fine channel fitting uncertainty is equal in amplitude to the thermal noise for the full experiment}. Hence, the thermal noise power expectation for the experiment is increased by a factor of two when fitting each fine channel independently, \textit{thereby reducing the overall array sensitivity by a factor of two}.

\subsection{Residual spectral power - amplitude residuals}
The fitting of a low-order polynomial (order $n$) over three coarse channels leaves residual spectral signal of, at least, order $n+1$ in each coarse channel. This residual signal is correlated across fine frequency channels, and hence generates residual power, with structure in the power spectrum parameter space. It is also subject to Runge's Effect, which occurs when fitting polynomials to regularly-discretized data  (see Section \ref{sec:results}). %The amplitude of this signal sets the tolerance level to bandpass calibration.

For each coarse channel for each of the real and imaginary components, we assume there remains a residual signal with unity mean (relative bandpass):
\begin{equation}
S_{r,i,{\rm res}}(\nu) = S_{\rm mn}\left(\frac{\nu-\nu_{c} - \xi_{r,i}}{\Delta\nu_{\rm coarse}}\right)^{n+1},
\label{eqn:resid_signal}
\end{equation}
where $\nu_{c}$ and $\Delta\nu_{\rm coarse}$ are the frequency offset and coarse channel bandwidth, respectively, and $S_{\rm mn}$ is a normalisation that sets the relative bandpass amplitude across the channel to equal unity. The offset $\xi_{r,i}$ translates the real and imaginary component residuals with respect to each other, and is a feature of a dipole antenna where the real and imaginary components are frequency-dependent. The mismatch occurs when combining voltages from antennas with slightly different characteristics (e.g., due to wind load or temperature), such that a sharp feature arises where there is a rapid change in the voltage response. This form allows a spectral feature in the bandpass amplitude at $\nu_c$, with an associated phase gradient with characteristic width $|\xi_r-\xi_i|$ (see Section \ref{sec:phase}). The residual signal, equation \ref{eqn:resid_signal}, is used to define the intra-channel correlation matrix, $\rho(\nu,\nu^\prime)$ (channels within each coarse channel are correlated, while inter-channel correlations are zero). This matrix is used, along with the sky point source signal model according to equation \ref{eqn:rho}, to define the power in the 2D power spectrum due to the fractional residual signal in the bandpass.

We consider an optimistic and a pessimistic model to describe the temporal and inter-station correlation of residual spectral curvature:
\begin{itemize}
\item Optimistic model: the most optimistic model for the residual signal is when each station's bandpass shape is different, and varies on timescales smaller than the calibration timescale. This leads to uncorrelated errors between stations and over time, yielding a stochastic signal that reduces with time;
\item Pessimistic model: the pessimistic model assumes that the station bandpass spectral shapes are the same across all stations (as described by the model dipole bandpass), and vary slowly in time. This model assumes complete correlation of residual errors between all stations and over a full 4-hour nightly track, thereby increasing the residual signal power by a factor of $(N_{\rm ant} \times t_{\rm track}/t_{\rm calibration})$ compared with the optimistic scenario.
\end{itemize}

In order to set the tolerance level, $\delta$, for the $n+1$-order polynomial residual, we scale the pessimistic model residual power until the amplitude of the thermal noise exceeds the residual across the full range of angular wavenumbers, $k_\bot$. \textit{This defines the maximum fractional bandpass error over a coarse channel}.

\subsection{Residual spectral power - phase residuals}\label{sec:phase}
Fitting of a $n$th-order polynomial to the real and imaginary components of the complex gain leaves at least ($n$+1)th-order residuals in both components. For the 3rd-order fit, the mismatch between the real and imaginary components yields a residual phase, according to:
\begin{equation}
\phi(\nu) = {\rm atan}{\frac{A_i}{A_r}\frac{(\nu-\nu_i-\xi/2)^4}{(\nu-\nu_r + \xi/2)^4}},
\end{equation}
where, $\xi = \xi_r - \xi_i$, and in the simplest case of equal amplitudes and turning points ($A_r=A_i$, $\nu_i=\nu_r$), has the form:
\begin{equation}
\phi(\nu) = {\rm atan}{\frac{(\nu-\nu_i-\xi/2)^4}{(\nu-\nu_i + \xi/2)^4}}.
\label{eqn:phase}
\end{equation}
Here, $\xi$ is the frequency mismatch between the real and imaginary components. This form encapsulates the features of a dipole antenna where there is a frequency-dependent change in the real and imaginary components of the voltage response. Such features are observed in early testing of the SKALA antenna \citep{delera15}.
% responses of the two polarizations are slightly mismatched (for the cross-polarization), and for single-polarization cross-correlation between non-identical dipoles.

\begin{figure*}[t]
{ \includegraphics[width=.5\textwidth]{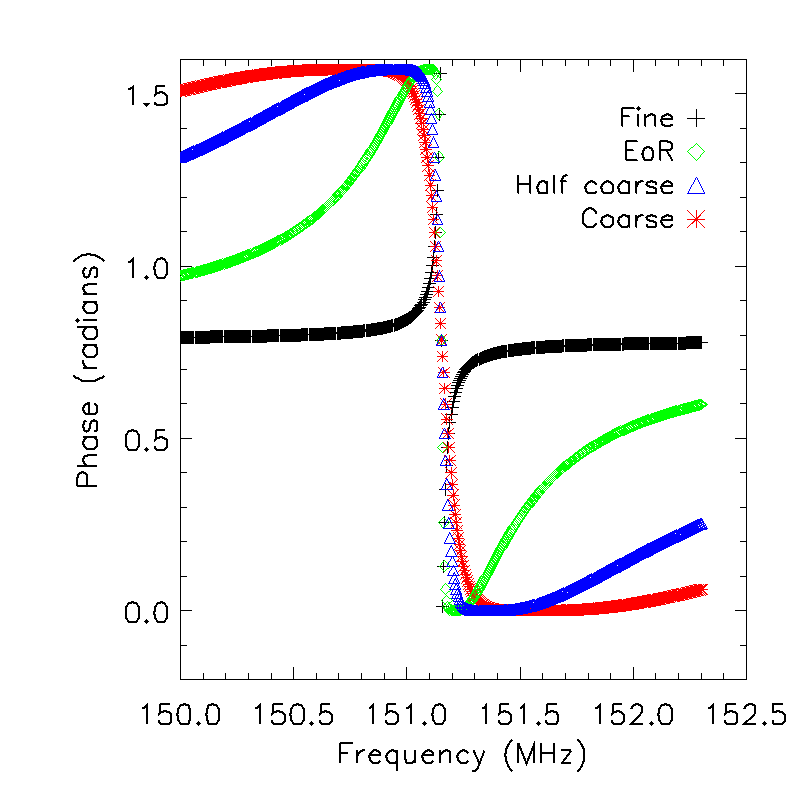}}
{ \includegraphics[width=.5\textwidth]{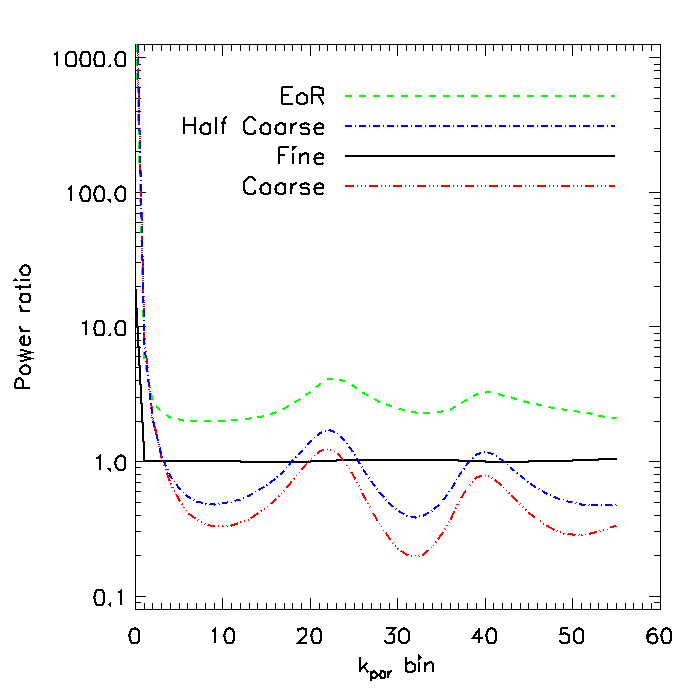}}
\caption{(Left) Phase plots as a function of frequency using equation \ref{eqn:phase}, where the mismatch in frequency is equivalent to a single fine channel (4.578 kHz; `Fine'), 24 fine channels (an EoR spectral channel; 109.87 kHz; `EoR'), one-half of a coarse channel (384.6 kHz; `Half coarse'), and a single coarse channel (769.10 kHz; `Coarse'). (Right) Ratio of power to a flat phase profile,  $\frac{|P_\xi|}{P_{\rm flat}}$, for a single value of $k_\bot$, and four values of the frequency mismatch, $\xi$.}
\label{fig:phase}
\end{figure*}

Figure \ref{fig:phase} (left) shows four representative phase plots as a function of frequency using equation \ref{eqn:phase}, where the mismatch in frequency is equivalent to a single fine channel ($\xi$=4.578 kHz; `Fine'), 24 fine channels (an EoR spectral channel; $\xi$=109.87 kHz; `EoR'), one-half of a coarse channel ($\xi$=384.6 kHz; `Half coarse'), and a single coarse channel ($\xi$=769.10 kHz; `Coarse'). Each symbol denotes a single fine channel.
The smaller the mismatch $\xi$, the steeper the phase feature. However, when the data are averaged to the EoR spectral resolution, mismatches finer than the channel resolution, 109.87 kHz, are smoothed.

We average each fine channel to the EoR spectral resolution, before forming the frequency-frequency correlation matrix, and propagating into the power spectrum. The errors are computed for a single coarse channel, with the remainder of the band showing no phase residuals. This therefore corresponds to the best-case scenario.

In phase, the correlation matrix contains complex phase terms that imprint phase structure. Figure \ref{fig:phase} (right) plots the ratio of power for each value of $\xi$, compared with a flat profile (zero phase), $\frac{|P_\xi|}{P_{\rm flat}}$, for a chosen cut through the 2D parameter space. The different phase structure imprints features at harmonics of the coarse bandpass, convolved with the spectral shape of the phase error. The phase errors \textit{bias} the cosmological signal by re-distributing power in LOS modes, but do not contribute power because the power spectrum natively destroys phase information, by squaring the complex visibilities.

\subsection{Results}\label{sec:results}
Figures \ref{fig:power_in50}, \ref{fig:power_in150} and \ref{fig:power_in200} display the 2D power spectra for each frequency, for: (top-left) The reference unresolved point source model; (top-centre) the calibration uncertainty power for a third-order polynomial fit; (top-right) the residual unresolved point source power due to residual fourth-order curvature (pessimistic model, scaled to reach thermal noise level); (bottom-left) the residual unresolved point source power due to residual fourth-order curvature (optimistic model); (bottom-centre) the uncertainty power due to calibration of each fine channel independently; and (bottom-right) the thermal noise power.

The strategy to calibrate each fine channel independently leads to no residual curvature (above the fine channel level) and no correlations, but is less precise than estimating the polynomial parameters because more independent parameters needs to be fitted, and therefore there are fewer available degrees of freedom. For bandpass responses that are smooth, this strategy costs a larger amount than the polynomial fitting, in the optimistic case.

\begin{figure*}[t]
\centering
\includegraphics[width=.95\textwidth]{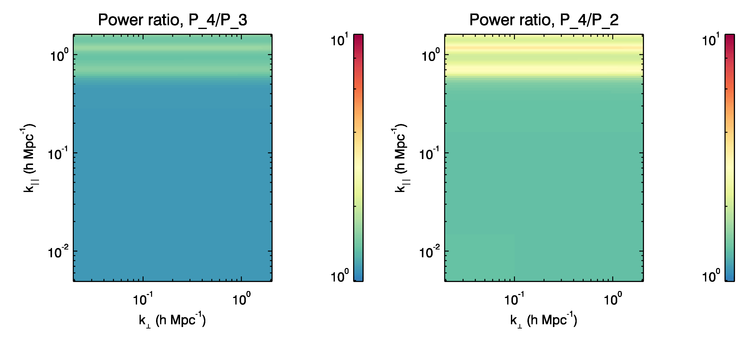}
\caption{Ratio of residual power after polynomial fitting, where the ratios compare a 4th-order fit to a 3rd-order fit (left), and a 4th-order fit to a 2nd-order fit (right). Runge's phenomenon is visible at high $k_\parallel$ where fitting residuals at the box edges are larger for higher order fits.}
\label{fig:runge_ratio}
\end{figure*}

In each experiment, the calibration uncertainty power due to fitting a third-order polynomial over three contiguous coarse channels yields structured power in the power spectrum space. This power is less than the thermal noise, but has the potential to cause low-level bias in the derived science results due to its shape. This is particularly true at low frequencies where the calibration uncertainty power is comparable to the thermal noise. The regular, horizontal lines of increased power in $k_\parallel$ are due to the correlation of uncertainties across a given coarse channel, but with independence between channels. This leads to a block-diagonal structure to the correlation matrix, and the features observed in the power spectrum. It is also a consequence of Runge's phenomenon \citep[e.g., see][]{dahlquist08}, whereby fitting a polynomial to regularly-space datapoints across a box leads to high-frequency residuals at the box edges. These residuals increase in amplitude for higher-order polynomial fits (for the same data spacing, as is the case here). This effect is visible in both the calibration uncertainty power (order $n$ fitting) and the residual power (order $n$+1 fitting). We can further demonstrate this by computing the ratio of residual power for different polynomial orders. Figure \ref{fig:runge_ratio} displays the ratio of the power in the 4th-order fit to the 2nd- and 3rd-order fits.
The additional power at high frequency (high $k_\parallel$) is visible for the higher-order fits.

This structure in $k_\parallel$ has the potential to yield bias in the results. As compared to the uncertainties (additional power) due to limited information, as computed by the Cramer-Rao bound, bias can add signal power that may be mistaken for cosmological power, in a statistical estimate. Despite potentially yielding a better fit to the data, the additional structured power from higher-order polynomials may be more problematic than using a fine-channel fitting procedure, where the residual power is higher, but is flat across the parameter space. This choice will be individual to the experiment itself, whereby the power at the location of the structures in $k_\parallel$ relative to the cosmological signal power, will determine the best approach (i.e., if the cosmological signal is expected to peak in the same region where the residual power gradient is large, then the bias will be increased). This is model, bandwidth and redshift-dependent for a given telescope.

Table \ref{table:tolerances} lists the relative fractional bandpass error tolerable in a ($n$+1)th-order polynomial residual, as applied to the plots, where $\delta$ is the fractional reduction in power values.
\begin{table}[h]
\centering
\begin{tabular}{|l|l|l|l|}
\hline 
Experiment & $\delta_{n=2}$ & $\delta_{n=3}$ & $\delta_{n=4}$\\
\hline
50~MHz & 0.027 & 0.025 & 0.019\\
100~MHz & 0.011 & 0.010 & 0.008\\
150~MHz & 0.006 & 0.005 & 0.004\\
200~MHz & 0.009 & 0.008 & 0.006\\
\hline
\end{tabular} 
\caption{Derived tolerances for each experiment such that the residual power due to ($n$+1)th-order curvature in the bandpass is less than the thermal noise.}
\label{table:tolerances}
\end{table}
Each tolerance is approximately one percent, but higher-order fits have more stringent constraints due to the additional structured power at high $k_\parallel$. Note that these results are appropriate when the underlying shape is well-fitted by an $(n+1)$th-order polynomial, such that the amplitude of terms with $>(n+1)$th-order are small compared with the $(n+1)$th order.

\begin{figure*}[t]
\centering
\subfigure[Fine]{
\includegraphics[width=.45\textwidth]{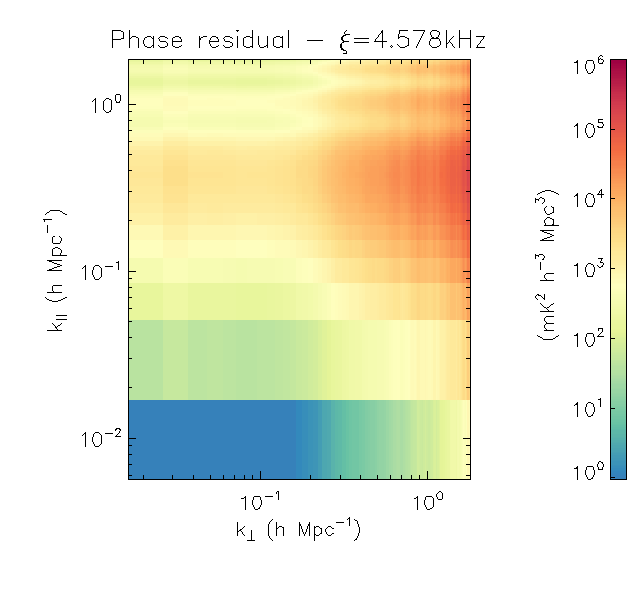}
}
\hspace{-2mm}
\subfigure[EoR]{
\includegraphics[width=.45\textwidth]{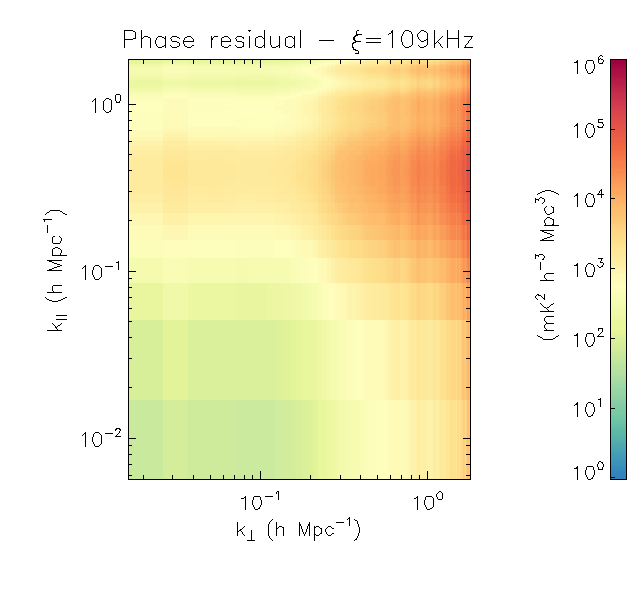}
}\\
\subfigure[Half a coarse channel]{
\includegraphics[width=.45\textwidth]{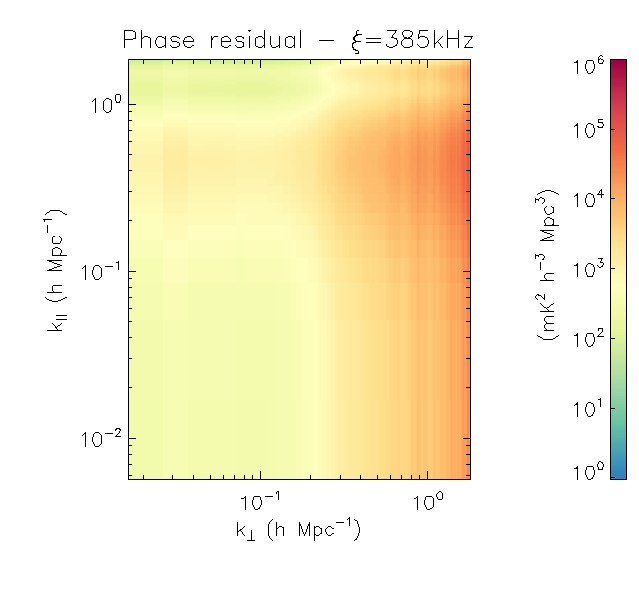}
}
\hspace{-2mm}
\subfigure[Coarse]{
\includegraphics[width=.45\textwidth]{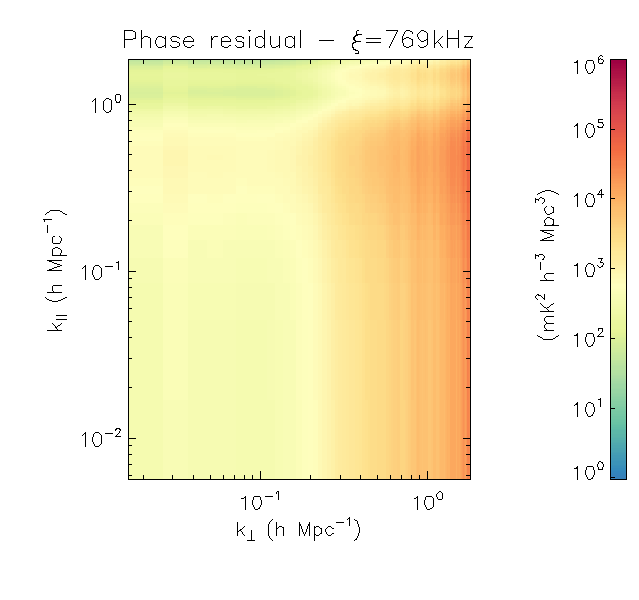}
}\\
\caption{2D power from phase residuals on sky point sources at 200~MHz, scaled by the amplitude tolerance, $\delta$=0.008, for four values of the frequency mismatch, $\xi$. The overall scale of power is not markedly changed, but the power is distributed differently in $k_\parallel$.}
\label{fig:phase_power}
\end{figure*}

The impact of phase residuals described in Section \ref{sec:phase} are plotted in the 2D power spectrum in Figure \ref{fig:phase_power}. Here, the same fractional reduction in power values, $\delta$, as described in Table \ref{table:tolerances} is applied to the sky point source model, in order to scale the uncertainty power to that deemed tolerable from amplitude residuals.%, \textit{but the amplitude error shape has been removed}, in order to display the phase residual component to the power. %Also, {\textit the phase residual is applied to only a single coarse band across the bandpass, with other coarse channels containing no residual}.
%\begin{figure*}[t]
%\centering
%\subfigure[Fine]{
%\includegraphics[width=.45\textwidth]{phase_2d_fine_200mhz.png}
%}
%\hspace{-2mm}
%\subfigure[Fine x 5]{
%\includegraphics[width=.45\textwidth]{phase_2d_eor_200mhz.png}
%}\\
%\subfigure[Fine x 24]{
%\includegraphics[width=.45\textwidth]{phase_2d_coarse_200mhz.png}
%}
%\hspace{-2mm}
%\subfigure[Coarse]{
%\includegraphics[width=.45\textwidth]{phase_2d_coarse3_200mhz.png}
%}\\
%\caption{Fractional differential power (equation \ref{eqn:phase}) compared with a smooth phase solution, for each value of the frequency mismatch, $\xi$.}
%\label{fig:phase_diff}
%\end{figure*}
Frequency mismatches below the EoR channel averaging bandpass (109.8~kHz) are smoothed, while much larger mismatches imprint a slower gradient structure. The largest differential is obtained for a frequency mismatch of the same scale as the spectral resolution (4.578~kHz), while the power distribution is most different from flat for a mismatch corresponding to the coarse channel bandwidth.

\begin{figure*}[t]
\centering
\subfigure[50~MHz.]{
\includegraphics[width=.45\textwidth]{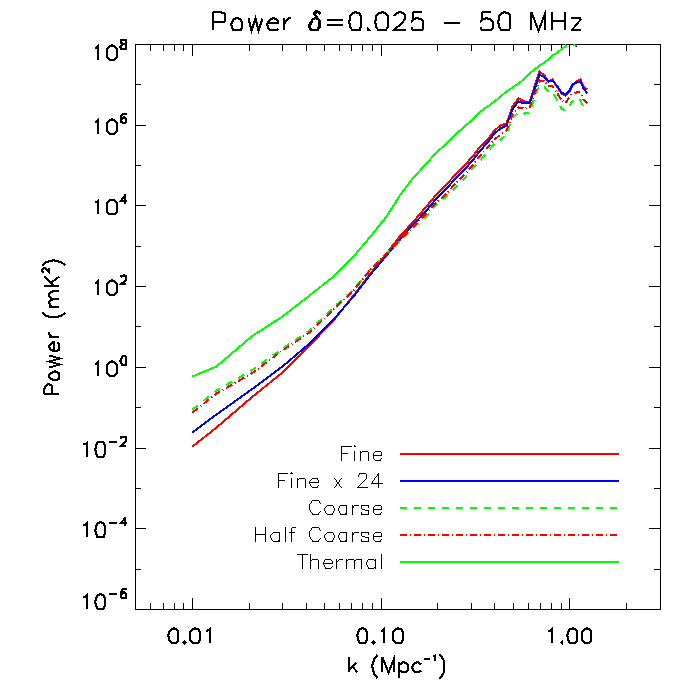}
}
\hspace{-2mm}
\subfigure[150~MHz]{
\includegraphics[width=.45\textwidth]{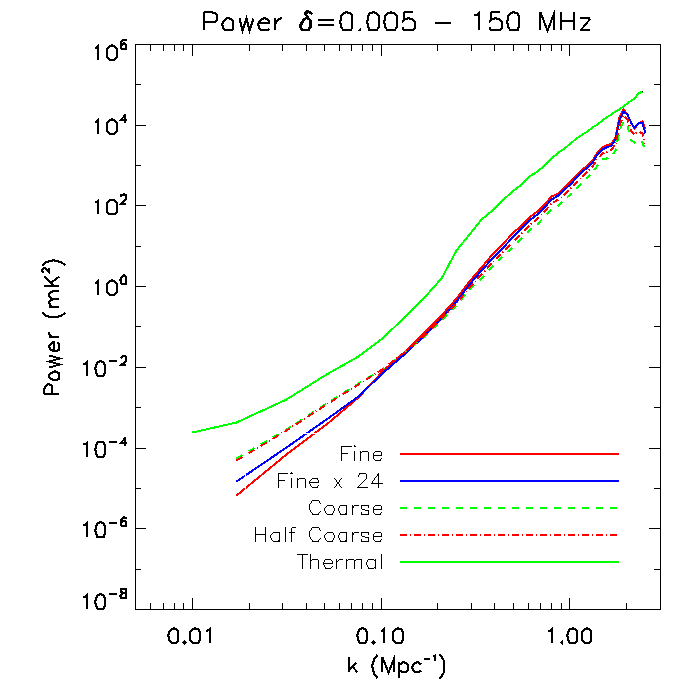}
}\\
\subfigure[200~MHz]{
\includegraphics[width=.45\textwidth]{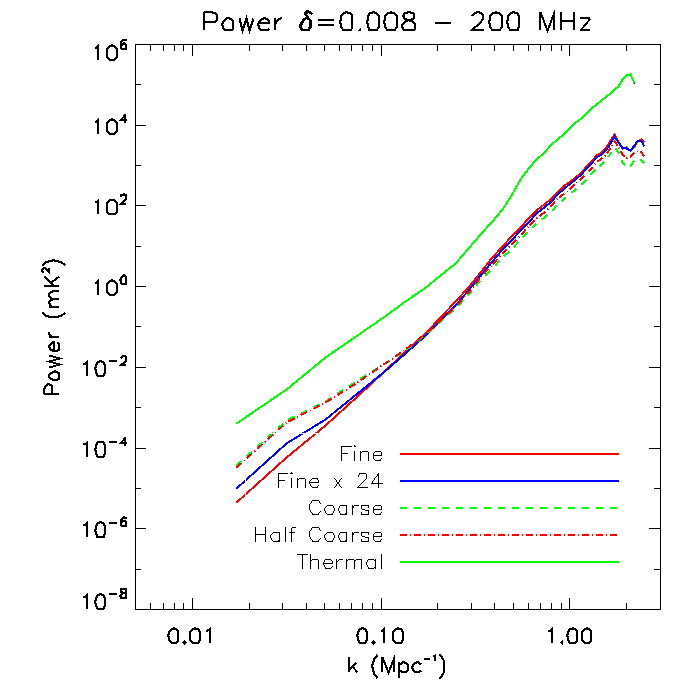}
}
\\
\caption{1D (spherically-averaged) power spectra for each value of the frequency mismatch, $\xi$, displaying the power bias introduced by a given combination of $\delta$ and $\xi$. Also shown is the thermal noise level.}
\label{fig:1d_phase}
\end{figure*}

When the amplitude and phase residuals are combined, the phase errors introduce a small power bias difference in the low $k$-modes. Figure \ref{fig:1d_phase} displays the spherically-averaged (1D) power from the amplitude and phase residuals, for each of four different values for $\xi$.
Smoother transitions in the spectrum (larger $\xi$) yield larger bias at low $k$, with factors of two reduction in power. There is, therefore, a bias in the power. The phase gradient tolerance derived from this work is weak, due to the implicit loss of phase information in the power spectrum upon squaring of the complex-valued data. A much more stringent gradient is set by the tomographic experiment, where phase is retained, and this is explored in the next section. As was shown for the 2D case in Figure \ref{fig:runge_ratio}, the structures at high $k_\parallel$ yield biases at large $k$.

\section{Tomography}
The tomographic experiment is a new addition to the EoR observational arsenal, not previously accessible with lower-sensitivity instruments. It is a challenging experiment that uses deep image cubes to detect brightness temperature fluctuations corresponding to ionised structures (e.g., bubbles) in the Cosmic Dawn and EoR. It is therefore a direct detection experiment, rather than a statistical experiment, necessitating high image dynamic range and precision calibration. We assess phase residual impact at 150~MHz, and at a wavenumber, $k$, where the array filling factor is close to unity.

There are several approaches to assessing the image dynamic range due to phase residuals, with a thorough treatment and rules-of-thumb presented in \citet{perley99}. Here, we wish to consider the impact of a residual phase gradient across a 100~kHz channel, with uncorrelated gradients between stations. To perform this analysis, we consider the maximum phase change across a fine channel. We define the phase within a fine channel for station $\alpha$ to be:
\begin{equation}
\phi_\alpha(\nu) = \epsilon_\alpha\nu,
\end{equation}
with a visibility at $\nu$ measured between two stations of:
\begin{equation}
V^\prime_{\alpha\beta}(\nu) = V_{\alpha\beta}(\nu)\exp{\left(-2\pi{i}(\epsilon_\alpha-\epsilon_\beta)\nu  \right)}.
\end{equation}
Integration of a signal across a channel leads to signal decorrelation according to the integral of this expression, such that the visibility degradation is given by:
\begin{equation}
\frac{V^\prime}{V} = \frac{1}{2\pi\Delta\epsilon\Delta\nu}\left[\sin{(2\pi\Delta\epsilon\Delta\nu)} + i(\cos{(2\pi\Delta\epsilon\Delta\nu)}-1) \right],
\label{eqn:dynamic_range}
\end{equation}
where $\Delta\epsilon$ and $\Delta\nu$ are the phase gradient differences between stations and channel width, respectively. One can immediately observe that the real part of this expression is a sinc function, which is $<$1 when $\Delta\epsilon\neq{0}$. This expression describes the fractional flux density loss of a coherent signal due to decorrelation across the bandwidth of a channel $\Delta\nu$.

We can now extend this analysis for a single visibility to explore the impact in an image slice, by defining the phase gradient for each station to be a random value, which is Gaussian-distributed with characteristic width $\sigma_\epsilon=\epsilon$. We proceed by forming the $uv$-sampling in a snapshot at zenith, with unity-amplitude visibilities corrupted according to equation \ref{eqn:dynamic_range}. We then Fourier Transform these gridded, corrupted visibilities to the image plane, and compare the peak of the point spread function (PSF) to that for a gridded set of unity-amplitude visibilities (i.e., the instrument sampling function, leading to the instrumental PSF). We define the tolerance to be the characteristic phase gradient, $\sigma_\epsilon$, where the reduction in dynamic range causes the residual from the brightest image source to exceed the expected bubble signal strength.
%Figure \ref{fig:decorrelation} shows the decorrelation as a function of phase residual across a channel ($\epsilon\Delta\nu$), characterised as the fractional loss in flux density of a source.
%\begin{figure}[t]
%\centering
%\includegraphics[width=.5\textwidth]{decorrelation.png}
%\caption{Reduction in flux density of a source due to decorrelation of signal across a frequency channel of width $\Delta\nu$, measured in radians. The phase residual gradient, $\epsilon$ sets the rate of decorrelation.}
%\label{fig:decorrelation}
%\end{figure}
A phase residual of 1 radian across a channel leads to almost complete decorrelation of signal.

\subsection{Results}
All sampled $uv$ points in the core are used to estimate the snapshot PSF of the array, matching the expected bubble size to the array resolution. At 150~MHz, the longer baselines provide resolution at the 10s arcseconds level, while the densely-packed core provides a wider PSF base of $\sim$10 arcmins. We compute the characteristic phase gradient, $\sigma_\epsilon$ on each station bandpass fine channel such that the dynamic range degradation destroys the cosmological signal. We take a pessimistic approach (where the phase residual on each channel is fixed over the full experiment) and an optimistic approach (where the phase residual is uncorrelated in time between calibration cycles, yielding an increase in dynamic range and weaker constraints). 

On scales of the SKA1-Low core (20 arcmin), a 1~mK brightness temperature fluctuation corresponds to a $\sim$0.02~mJy source, increasing to 0.2~mJy for degree scales \citep{koopmans15}. For a 5~Jy source in the field, these levels corresponds to a 0.02/5000=10$^{-5}$ (0.2/5000=10$^{-4}$) fractional residual flux density (noise term in the image) across the array in a 4.58~kHz channel for the pessimistic case. Loss in dynamic range of this amplitude would erase any high-redshift signal, and sets the tolerance level. Any 1~degree bubble is statistically likely to have a 1~Jy source within its synthesized beam, and will therefore be affected by this effect. Table \ref{table:decorrelation} shows the derived parameters for each experiment.
\begin{table}[h]
\centering
\begin{tabular}{|c|c|c|}
\hline 
Scale & $\sigma_\epsilon\Delta\nu$ (degrees) & $\sigma_\epsilon\Delta\nu$ (degrees)\\ 
 & Pessimistic & Optimistic\\
\hline 
20' & 0.04 & 0.2 \\ 
60'=1 deg. & 0.2 & 1.3\\ 
\hline 
\end{tabular} 
\caption{Characteristic phase residual across a fine channel, $\sigma_\epsilon\Delta\nu$, for a loss in dynamic range that would destroy the cosmological signal, due to phase decorrelation.}
\label{table:decorrelation}
\end{table}
The constraints are therefore stringent in either the pessimistic or optimistic case for bubbles on tens of arcminute scales.

The tolerance allows for a phase gradient that yields a phase difference that does not exceed a residual of 0.2 (1.3) degrees per fine channel in the pessimistic (optimistic) case, on scales of 1 degree. This calculation assumes that the phase residual is uncorrelated between two different antennas.

\section{Summary}
This work takes a signal estimation theoretic approach to calibration of low-frequency radio interferometers. Specifically, it computes the tolerances on bandpass amplitude and phase residuals for SKA-Low to achieve the goals of the Epoch of Reionisation and Cosmic Dawn experiments, using calibration precision and accuracy. The additional power due to imprecise and inaccurate bandpass calibration in the power spectrum is defined to not exceed the thermal noise power expected for the experiment. Similarly, the imaging dynamic range is defined to not destroy brightness temperature fluctuation detections for the tomography experiment. These tolerances provide quantitative specifications for the intrinsic spectral response of SKA-Low antennas. The choice of calibration approach depends on the shape of the antenna bandpass, whereby the requirement to fit higher-order polynomials may lead to increased signal power bias. In this case, a fine-channel fitting model, while yielding larger uncertainties across the parameter space, also yields a smooth response in frequency, thereby producing a potentially less-biased solution.

\section*{Acknowledgements}
This research was supported under Australian Research Council's Discovery Early Career Researcher funding scheme (project number DE140100316) and the Centre for All-sky Astrophysics (an Australian Research Council Centre of Excellence funded by grant CE110001020). This work was supported by resources provided by the Pawsey Supercomputing Centre with funding from the Australian Government and the Government of Western Australia. We acknowledge the International Centre for Radio Astronomy Research (ICRAR), a Joint Venture of Curtin University and The University of Western Australia, funded by the Western Australian State government. The authors would like to thank Peter Dewdney, Jeff Wagg and Eloy de Lera Acedo for useful discussions.

%\nocite*{}
\bibliographystyle{jphysicsB}

\bibliography{skabib.bib}

%suggested order:\\
%- intro to the experiments\\
%- intro to the aperture arrays\\
%- discussion of calibration: implicit model, dynamic range limits, SKA memos so far \\
%- discussion of what's special about the power spectrum experiment, why bandpass cal matters so much \\
%- intro to SKALA, motivation for paper

\end{document}